\documentclass{desyproc}
\usepackage{amsmath}
\usepackage{amssymb}
\usepackage{fancyhdr}

\newcommand{\vect}[1]{\mathbf{#1}}

\begin{document}


\title{Single and Double Perturbative Splitting \\ Diagrams in Double Parton Scattering}



%
%
%
%
%
%
%
%
%

\author{{\slshape Jonathan R. Gaunt$^1$\footnote{Speaker}, W. James Stirling$^1$}\\[1ex]
     $^1$Cavendish Laboratory, J J Thomson Avenue, Cambridge CB3 0HE, UK}


\contribID{ZZ}
\confID{UU}
\desyproc{DESY-PROC-2012-YY}
\acronym{MPI@LHC 2011}
\doi
\maketitle


\begin{abstract}

We discuss the role of two different types of diagram in the proton-proton double parton scattering (DPS) cross 
section -- single and double perturbative splitting graphs. Using explicit calculations 
of simple graphs from these classes we show that the treatment of these graphs by the
`double PDF' framework for describing the DPS cross section, introduced a number of years
ago by Snigirev and collaborators, is unsatisfactory. We suggest that a contribution from 
single perturbative splitting graphs should be included in the DPS cross section, albeit
with a different geometrical prefactor to the contribution from `zero perturbative splitting'
graphs.

\end{abstract}



\section{`Double Perturbative Splitting' Diagrams in Double Parton Scattering} \label{sec:splittingdiags}

\thispagestyle{fancy}

We define double parton scattering (DPS) as the process in which two pairs of partons participate in hard interactions in a 
single proton-proton (p-p) collision. DPS processes can constitute important backgrounds to Higgs and other interesting signals (see 
e.g. \cite{DelFabbro:1999tf}), and can themselves be considered as interesting signal processes, since they reveal information 
about parton pair correlations in the proton.

Making the assumption that the hard processes A and B may be factorised, the cross section for p-p DPS 
may be written as follows:
\begin{align} \label{DPSXsec1}
\sigma^D_{(A,B)} \propto &\sum_{i,j,k,l}\int \prod_{a=1}^{4}dx_a d^2\vect{b}
\hat{\sigma}_{ik \to A}(\hat{s} = x_1x_3s) \hat{\sigma}_{jl \to B}(\hat{s} = x_2x_4s) \\ \nonumber
&\times \Gamma_{ij}(x_1,x_2,\vect{b};Q_A^2,Q_B^2)\Gamma_{kl}(x_3,x_4,\vect{b};Q_A^2,Q_B^2) 
\end{align}

The cross section formula is somewhat similar to that used for single parton scattering (SPS), except that two 
parton-level cross sections $\hat{\sigma}$ appear, and the PDF factors are two-parton generalised PDFs $\Gamma$
(2pGPDs) rather than single PDFs. Note that in this formula the two 2pGPDs are integrated over a common parton 
pair transverse separation $\vect{b}$.

In many extant studies of DPS, it is assumed that the 2pGPD can be approximately factorised into a product of a
longitudinal piece and a (typically flavour and scale independent) transverse piece:
\begin{equation} \label{2pGPDdecomp2dPDF}
\Gamma_{ij}(x_1,x_2,\vect{b};Q_A^2,Q_B^2) \simeq D_p^{ij}(x_1,x_2;Q_A^2,Q_B^2) F(\vect{b})
\end{equation}

Then, if one introduces the quantity $\sigma_{\mathrm{eff}}$ via $\sigma_{\mathrm{eff}} \equiv 1/[\int F(\vect{b})^2 d^2 \vect{b}]$, one 
finds that one may write $\sigma^D_{(A,B)}$ entirely in terms of the longitudinal piece and $\sigma_{\mathrm{eff}}$:
\begin{align} \label{dPDFXsec}
\sigma^{D}_{(A,B)} \propto& \dfrac{1}{\sigma_{\mathrm{eff}}}\sum_{i,j,k,l}\int \prod_{a=1}^{4}dx_a D_p^{ij}(x_1,x_2;Q_A^2,Q_B^2) 
D_p^{kl}(x_3,x_4;Q_A^2,Q_B^2)
 \hat{\sigma}_{ik \to A} \hat{\sigma}_{jl \to B}
\end{align}

In \cite{Zinovev:1982be} a quantity $D_p^{ij}(x_1,x_2; Q^2)$ is introduced, and an evolution equation for this quantity is
given. We shall refer to the quantity and its evolution equation as the double PDF (dPDF) and the dDGLAP equation respectively. It is
asserted in \cite{Snigirev:2003cq} that the dPDF is equal to the factorised longitudinal part of the 2pGPD in the case in
which the two hard scales $Q_A^2$ and $Q_B^2$ are equal to a common value $Q^2$. 

\begin{figure}
\centering
\includegraphics[scale=0.5, trim = 0 1cm 0 2cm]{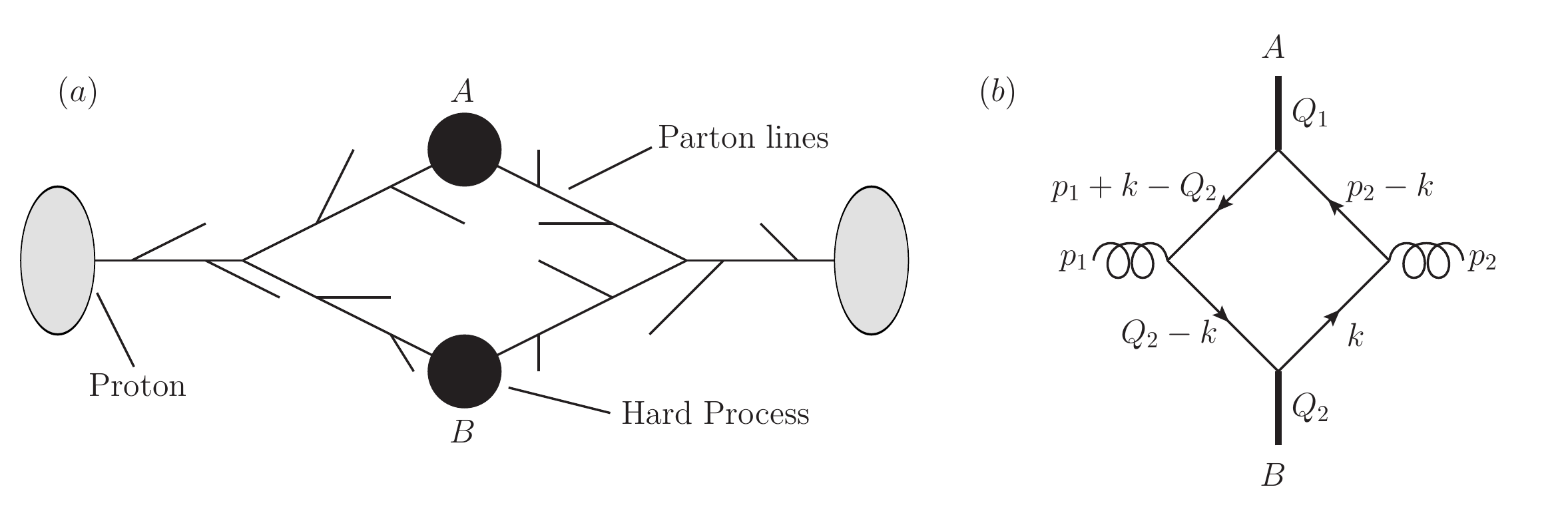}
\caption{\label{fig:dpsloops} (a) A diagram that apparently contributes to the leading order DPS cross section according to the
framework of \cite{Snigirev:2003cq}. The partons emerging from the grey proton blobs in the figure are nonperturbatively generated
partons -- i.e. ones existing at a low scale $\sim \Lambda_{QCD}$.
(b) The `crossed box' graph. In this part of the figure, A and B are arbitrary single particle final states with 
$Q_1^2 = Q_2^2 = Q^2 > 0$.}
\end{figure}

The dDGLAP equation contains two types of terms on the right hand side -- `independent branching' terms corresponding to
emission of partons from a pre-existing pair, and `single parton feed' terms corresponding to the perturbative generation
of a pair from the splitting of a single parton. The single feed terms involve the leading twist single parton 
distributions as one might expect. Given this structure of the dDGLAP equation, with single feed terms included on the
right hand side, a prediction of the `dPDF framework' suggested in \cite{Snigirev:2003cq} for calculating the p-p DPS
cross section is that a part of the `double perturbative splitting' or `1v1' graph drawn in figure \ref{fig:dpsloops}(a) 
should be included in the LO p-p DPS cross section. The part that should be included is proportional to 
$[\log(Q^2/\Lambda^2)]^n/\sigma_{\mathrm{eff}}$ at the cross section level, where $\Lambda$ is some IR cutoff of order 
$\Lambda_{QCD}$, and $n$ is equal to the total number of QCD branchings in figure \ref{fig:dpsloops}(a) (including the 
two that only produce internal particles). This piece should be associated with the region of transverse momentum integration 
for the graph in which the transverse momenta of the branchings on either side of the `hard processes' in the graph are 
strongly ordered up to scale $\sim Q^2$.

The question that then arises is whether such a structure in fact exists in the cross section expression for the loop of figure 
\ref{fig:dpsloops}(a). Starting from the conventional `Feynman rules' expression for the loop, it is not immediately obvious 
what the answer to this question is. Here we will focus on answering this question for the specific very simple `crossed box' 
loop shown in figure \ref{fig:dpsloops}(b), which is predicted by the dPDF framework to contain a piece proportional to 
$[\log(Q^2/\Lambda^2)]^2/\sigma_{\mathrm{eff}}$\footnote{Note that the `crossed box' topology of figure \ref{fig:dpsloops}(b) is
the only box topology that contains a DPS singularity.}. The issues raised in the treatment of this example carry over to the more general 
loop of figure \ref{fig:dpsloops}(a). 

We expect the $[\log(Q^2/\Lambda^2)]^2/\sigma_{\mathrm{eff}}$ piece in figure \ref{fig:dpsloops}(b) to be predominantly contained in the
portion of the cross section integration in which the external transverse momenta, as well as the transverse momenta and virtualities
of the internal particles, are all small. This is actually the region around a certain pinch singularity in the loop integral known
as the double parton scattering singularity \cite{Nagy:2006xy}. In \cite{Gaunt:2011xd}, we obtained an analytic expression for the
part of an arbitrary loop containing a DPS singularity associated with the loop particles emerging from the initial state particles
being nearly on-shell and collinear, in the limit in which the external transverse momenta are small. Applied to the loop of figure
\ref{fig:dpsloops}(b) this reads (schematically, suppressing helicity and colour labels):
\begin{align} \label{DPSamp}
L_{DPS, \text{fig \ref{fig:dpsloops}(b)}} \propto \dfrac{1}{Q^2} & \int d^2 \vect{k}  \Phi_{g \to q\bar{q}} (x, \vect{k-Q_2})
 \Phi_{g \to \bar{q}q} (1-x, \vect{-k}) 
\\ \nonumber &\times 
\mathcal{M}_{q\bar{q} \to A} (\hat{s} = x(1-x)s) \mathcal{M}_{q\bar{q} \to B} (\hat{s} = x(1-x)s) + (q \leftrightarrow \bar{q})
\end{align}

In this formula, $x = p_2 \cdot Q_1 / p_1 \cdot p_2$, $s = (p_1+p_2)^2$, and $\vect{k}$ ($\vect{Q_2}$) is the component of 
$k$ ($Q_2$) transverse to the axis defined by the directions of the incoming particles.  $\Phi_{g \to q\bar{q}}(x,\vect{k})$ 
is the $\mathcal{O}(\alpha_S)$ light cone wavefunction (LCWF) to produce a $q\bar{q}$ pair from a $g$ \cite{Harindranath:1998pd}, 
with the quark having lightcone momentum fraction $x$ and transverse momentum $\vect{k}$ with respect to the parent gluon. It can 
be factored into a $\vect{k}$ and $x$ dependent part, where the $\vect{k}$ dependent part is proportional to $\vect{\epsilon \cdot 
k}/\vect{k}^2$, $\vect{\epsilon}$ being the transverse part of the gluon polarisation vector. It is generally true that any QCD 
$1 \to 2$ splitting with physically polarised external particles has a corresponding LCWF that is proportional to $1/\vect{k}$.
Provided one uses a physical gauge for the gluon, all LCWFs corresponding to a $1 \to 2$ QCD splitting are then proportional to 
$1/\vect{k}$.

Inserting \eqref{DPSamp} into the standard $2 \to 2$ cross section expression, and performing a number of changes of variable, we
arrive at the following expression for the DPS singular part of the $gg \to AB$ cross section:
\vspace{-5mm}
\begin{align} \label{ggDPSXsec2}
\sigma_{DPS, \text{fig \ref{fig:dpsloops}(b)}} \propto & \int \prod_{i=1}^{2}dx_id\bar{x}_i \hat{\sigma}_{q\bar{q} \to A}(\hat{s} = x_1\bar{x}_1s) 
\hat{\sigma}_{q\bar{q} \to B} (\hat{s} = x_2\bar{x}_2s)
\\ \nonumber
\times & \int \dfrac{d^2\vect{r}}{(2\pi)^2} 
\Gamma_{g \to q\bar{q}}(x_1,x_2, \vect{r}) \Gamma_{g \to \bar{q}q}(\bar{x}_1,\bar{x}_2, -\vect{r}) 
\end{align}
\vspace{-5mm}
\begin{align} \label{rspace2pGPD}
\quad \Gamma_{g \to q\bar{q}}(x_1,x_2, \vect{r}) \propto \dfrac{\alpha_S}{2\pi} \delta(1-x_1-x_2) T^{ij} (x_1,x_2)
 \int^{\vect{\tilde{k}}^2 < \mathcal{O}(Q^2)} d^2 \vect{\tilde{k}} \tfrac{[\vect{\tilde{k}}+\tfrac{1}{2}\vect{r}]^i [\vect{\tilde{k}}-\tfrac{1}{2}\vect{r}]^j}
{[\vect{\tilde{k}}+\tfrac{1}{2}\vect{r}]^2 [\vect{\tilde{k}}-\tfrac{1}{2}\vect{r}]^2}.
\end{align} 

$T^{ij} (x_1,x_2)$ contains a function of $x_1$ and $x_2$ that may be regarded as a `$1 \to 2$' splitting function, 
multiplied by a constant matrix in transverse space\footnote{Note that the cross section is really a sum of terms with 
different $T^{ij} (x_1,x_2)$ factors in the $g \to q\bar{q}$ 2pGPDs. This is associated with the fact that, from the 
point of view of the quarks, there is an unpolarised diagonal contribution to the process plus polarised and interference
contributions in colour, spin, and flavour space. See e.g. \cite{Diehl:2011tt,Diehl:2011yj} for a discussion of 
correlation and interference effects in DPS processes.}. $\vect{r}$ is equal to the transverse momentum imbalance of one 
of the quarks/antiquarks in the loop between amplitude and conjugate, and is the Fourier conjugate variable of the parton pair 
separation $\vect{b}$ in the $q\bar{q}$ pair emerging from either gluon. $\Gamma_{g \to q\bar{q}}(x_1,x_2, \vect{r})$ 
can therefore be thought of as the $\mathcal{O}(\alpha_S)$ transverse momentum-space 2pGPD to find a $q\bar{q}$ pair 
inside a gluon. Note that the expression here effectively coincides with that of \cite{Diehl:2011tt}, in which a cross 
section expression for the box of \ref{fig:dpsloops}(b) is obtained starting from a pure DPS view of the box.

Let us consider the part of the integral \eqref{ggDPSXsec2} that is associated with the magnitude of the 
imbalance $\vect{r}$ being smaller than some small cut-off $\Lambda$ that is of the order
of $\Lambda_{QCD}$. The contribution to the cross section from this portion contains a $\log^2(Q^2/\Lambda^2)$ factor 
multiplied by $\Lambda^2$ (which can be thought of as an effective `$1/\sigma_{\mathrm{eff}}$' factor for this contribution). 
The majority of this contribution comes from the region in which the transverse momenta and virtualities of the 
quarks and antiquarks in the $gg \to AB$ loop are much smaller in magnitude than $\sqrt{Q^2}$ (i.e. the region in 
which the assumptions used to derive \eqref{DPSamp} apply), which is a necessary feature of a contribution to 
be able to regard it as a DPS-type contribution. By making a specific choice of $\Lambda$ (let us call this $\Lambda_S$), 
one could obtain an expression which is exactly in accord with the expectations of \cite{Snigirev:2003cq} -- that is, 
a product of two large DGLAP logarithms multiplied by the same $1/\sigma_{\mathrm{eff}}$ factor that appears in diagrams 
in which the parton pair from neither proton has arisen as a result of one parton perturbatively splitting into two
(`2v2' or `zero perturbative splitting' diagrams). The $1/\sigma_{\mathrm{eff}}$ factor for the 2v2 diagrams presumably has 
a natural value of the order of $1/R_p^2$ that is set by the nonperturbative dynamics ($R_p$ = proton radius).

The fact that we have to make a somewhat arbitrary choice for $\Lambda$ in order to arrive at the result anticipated
by the dPDF framework is concerning. There is nothing in the calculation of figure \ref{fig:dpsloops}(b) 
to indicate that we should take the region of it with $|\vect{r}| < \Lambda_S$ as the `DPS part' -- the scale $\Lambda_S$ does not 
naturally appear at any stage of the calculation. There is no more justification for taking the part of the box with
$|\vect{r}| < \Lambda_S$ to be the DPS part than there is for, say, taking the piece with $|\vect{r}| < 2\Lambda_S$,
or that with $|\vect{r}| < \Lambda_S/2$, to be the DPS part. We have had to artificially introduce the cut-off $\Lambda_S$ in 
figure \ref{fig:dpsloops}(b) in order to obtain a power-suppressed DPS part because there is no scale in the graph apart from 
$Q^2$, so in order to obtain a term proportional to $1/Q^2$, a second scale has to be introduced `by hand'\footnote{This is related to
the fact that in massless perturbation theory, there are no power corrections.}.

There therefore appear to be some unsatisfactory features of the dPDF framework with regards 
to its treatment of the box in figure \ref{fig:dpsloops}(b). In a physical gauge, precisely the same issues will 
be encountered for the case of the arbitrary `1v1' graph in figure \ref{fig:dpsloops}(a). 
One obtains a result that is consistent with the dPDF framework if one demarcates the portion 
of the cross section integral in which the transverse loop momentum imbalance between amplitude and conjugate is 
less than $\Lambda_S$ as DPS, but there is no natural reason to do this. There is no distinct piece of figure 
\ref{fig:dpsloops}(a) that contains a natural scale of order $\Lambda_{QCD}$ and is associated with the 
transverse momenta inside the loop being strongly ordered on either side of the diagram. In fact, most of 
the contribution to the total cross section expression for the graph comes from the region of integration in 
which the transverse momenta of particles inside the loop are of $\mathcal{O}(\sqrt{Q^2})$\footnote{One should 
bear in mind however that the same is not true for the cross section expression differential in the transverse 
momenta of $A$ and $B$ for $p_{T,A}^2, p_{T,B}^2 \ll Q^2$. Here, the major contribution is associated with 
transverse momenta in the loop $\ll \sqrt{Q^2}$ if there is one emission or more from inside the loop, or 
with a range of transverse momenta between $\sim |p_{T,A}|, |p_{T,B}|$ and $\sim \sqrt{Q^2}$ if there are no 
such emissions \cite{Diehl:2011yj}.}. This fact suggests that at the level of total cross sections, we should 
perhaps remove `1v1' graphs entirely from the DPS contribution, and regard them as 
pure SPS (this approach is advocated in \cite{Blok:2011bu}, for both the total and the differential cross 
sections). Treating the graphs in this way would have the advantage that we would not perform any double counting 
between DPS and SPS -- the graph of figure \ref{fig:dpsloops}(a) is in principle also included in the SPS 
$pp \to AB$ cross section (albeit as a very high order correction that will not be included in practical 
low order calculations, if the number of QCD emissions from inside the loop of the graph is large). 

Very similar conclusions may be reached if one uses a covariant gauge such as the Feynman gauge for the gluon
fields in figure \ref{fig:dpsloops}(a), although these conclusions are perhaps not obtained so readily. In a 
covariant gauge, gluons with unphysical `scalar' polarisation can exist in loop diagrams. Such scalar-polarised 
gluons can give rise to power-law DPS divergences rather than logarithmic ones, and additional `super-leading' 
contributions to the $AB$ production process (in terms of powers of $Q$) -- the two phenomena are related. On 
the other hand one generally expects the `super-leading' contribution to cancel in a suitable sum over graphs (as in 
\cite{Labastida:1984gy}), which effectively leaves one with the same logarithmic DPS divergences that are
encountered in a physical gauge.

One can gain some insight into the source of the problems in the dPDF framework by looking at
the $\vect{b}$-space 2pGPD corresponding to \eqref{rspace2pGPD}, $\tilde{\Gamma}_{g \to q\bar{q}}(x_1,x_2, \vect{b})$.
This comes out as being proportional to $1/\vect{b}^2$ -- this behaviour (which was first spotted in \cite{Diehl:2011tt}) 
can be traced to the fact that the $g \to q\bar{q}$ LCWF in $\vect{b}$ space (like any any LCWF corresponding to a QCD 
$1 \to 2$ splitting with physically polarised external particles) is proportional to $1/\vect{b}$, and $\tilde{\Gamma}(\vect{b}) 
\sim \Phi(\vect{b})^2$. Note that this behaviour is very different from the behaviour of all 2pGPDs that is anticipated by the 
dPDF framework (i.e. smooth function of size $R_p$). There is no natural feature in the product of two `perturbative splitting' 
2pGPDs that is of size $R_p$ and can be naturally identified as DPS. A key error then in the formulation of the dPDF framework 
seems to be the assumption that all 2pGPDs can be approximately factorised into dPDFs and smooth transverse functions of size 
$R_p$. A sound theoretical framework for describing p-p DPS needs to carefully take account of the different $\vect{b}$ 
dependence of pairs of partons emerging from perturbative splittings, whilst simultaneously avoiding double counting between 
SPS and DPS.

\vspace{-3mm}

\section{`Single Perturbative Splitting' Diagrams} \label{sec:2v1graphs}

\begin{figure}
\centering
\includegraphics[scale=0.5, trim = 0.5cm 0 0 2cm]{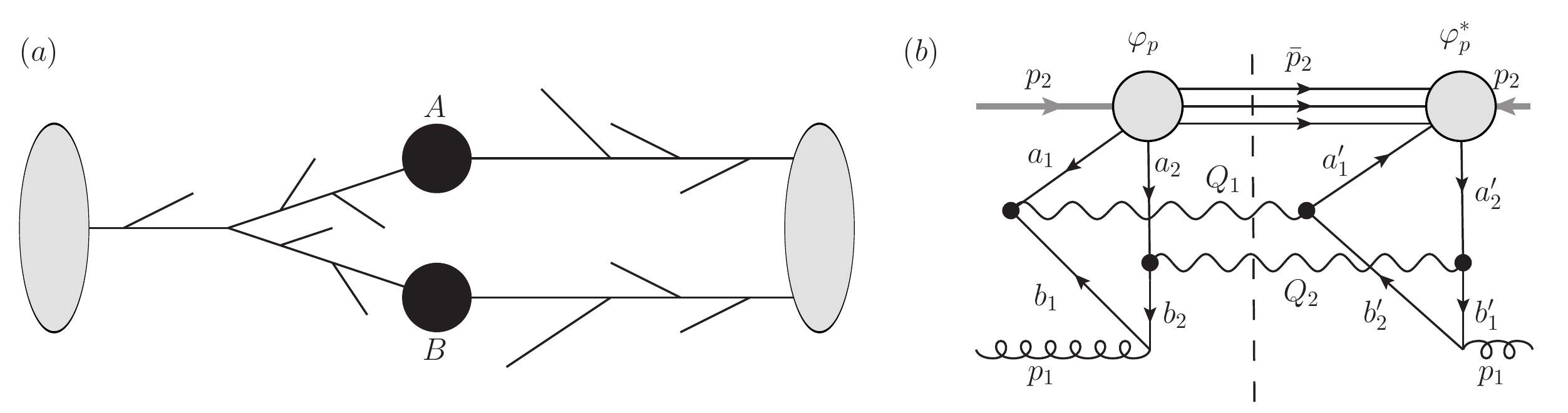}
\caption{\label{fig:2v1graphs} (a) A generic graph from the `2v1' class.
(b) A simple `2v1' graph in which a gluon splits into a $q\bar{q}$ pair, and these partons then
interact in two separate Drell-Yan interactions with a `nonperturbatively generated' $q\bar{q}$ pair 
from a proton.}
\end{figure}

Aside from the `1v1' graphs that were the focus of the previous section, and the `2v2' graphs that were also briefly 
mentioned, there is a further class of graph that can potentially contribute to the p-p DPS cross section.
These are graphs in which one proton provides one parton to the double scattering, and the other two, at the 
nonperturbative level -- a representative graph is sketched in figure \ref{fig:2v1graphs}(a). For obvious reasons, 
we will refer to the graphs as `2v1' or `single perturbative splitting' graphs.

It seems clear that we should include contributions from the 2v2 graphs as part of the DPS cross section.
An important question is whether we should also include contributions from the 2v1 graphs, and if so, what form
these contributions should take (in particular, how does the effective $\sigma_{\mathrm{eff}}$ factor for the 2v1 graphs
differ from that for the 2v2 graphs?).

To answer this question, let us take a similar approach as we did for the 1v1 graphs in the previous section.
That is, we take the graph drawn in figure \ref{fig:2v1graphs}(b) that has the simplest possible 2v1 structure,
and see whether there is a `natural' part of the cross section expression for it that is proportional to $1/R_p^2$, 
and also contains a large logarithm associated with the $1 \to 2$ splitting. If there is such a structure in the 
2v1 graph, then this part of this graph should be included in the LO DPS cross section, and we also expect there to
be a $\log(Q^2/\Lambda^2)^n/R_p^2$ piece in the more general 2v1 diagram of figure \ref{fig:2v1graphs}(a) that should 
also be included in the LO DPS cross section.

In the calculation of the cross section for figure \ref{fig:2v1graphs}(b), it is necessary to include a hadronic 
amplitude or wavefunction factor $\psi_p$ to find two nonperturbatively generated (`independent') partons in the 
proton, at the amplitude level in the calculation. It would be inappropriate to try and calculate a 2v1 cross section 
in a naive `fully parton-level' way omitting the proton on the `nonperturbative pair' side because then one has three particles 
in the initial state (whereas the standard framework for calculating a cross section requires two particles in the 
initial state). Furthermore, by deleting the proton on the `nonperturbative pair' side one is then neglecting the important
fact that the nonperturbatively generated partons are tied together in the same proton (as was pointed out in
\cite{Blok:2011bu}). The use of proton wavefunctions or hadronic amplitudes in the calculation of DPS-type graphs
was discussed long ago in \cite{Paver:1982yp}, and has been discussed more recently in \cite{Diehl:2011yj, Blok:2011bu}.

After a lengthy calculation, one finds that the cross section for figure \ref{fig:2v1graphs}(b) contains the 
following expression:
\begin{align} \label{2v1XS4}
\sigma_{1v2}(s) =& 
\hat{\sigma}_{\bar{q}q \to \gamma*}(\hat{s} = x_1y_1s)
\hat{\sigma}_{q\bar{q} \to \gamma*}(\hat{s} = x_2y_2s)
\\  \nonumber
& \times \dfrac{m}{2}\left[\int \dfrac{d^2\vect{r}}{(2\pi)^2} \Gamma_{p;q\bar{q}}\left(x_1,x_2;\vect{r}\right)\right]\left[\dfrac{\alpha_s}{2\pi}
P_{g \to q\bar{q}}\left(y_2\right)\delta(1-y_1-y_2) \int_{\Lambda^2}^{Q^2} \dfrac{d\vect{Q_1}^2}{\vect{Q_1}^2}\right]
\end{align}
In this expression we have omitted helicity and colour labels and sums for simplicity. The quantity 
$\Gamma_{p;q\bar{q}}\left(x_1,x_2;\vect{r}\right)$ is the 2pGPD of the nonperturbatively generated
parton pair, whilst $P_{g \to q\bar{q}}$ is the LO $1 \to 2$ splitting function
for the process $g \to q\bar{q}$. $m$ is a symmetry factor that is equal to $1$ if the two hard processes are the 
same, and is equal to $2$ otherwise (for the double Drell-Yan process under consideration, it equals $1$).

The integral over $\vect{Q_1}$ in \eqref{2v1XS4} gives rise to a large transverse momentum logarithm $\log(Q^2/\Lambda^2)$, 
whilst the integral over $\vect{r}$ supplies a prefactor of order $\Lambda^2 \sim 1/R_p^2$ (we assume $\Gamma_{p;q\bar{q}}
\left(\vect{r}\right)$ only has support for $|\vect{r}|$ values of order $\Lambda_{QCD}$ -- see later). Thus, there is a 
part of the cross section expression for figure \ref{fig:2v1graphs}(b) that is proportional to $\log(Q^2/\Lambda^2)/R_p^2$ 
and should be included in the LO DPS cross section.

When we generalise the result \eqref{2v1XS4} to the leading logarithmic part of the arbitrary 2v1 diagram in figure
\ref{fig:2v1graphs}(a), and then sum up all of the diagrams to obtain the contribution of 2v1 graphs to the LO DPS
cross section, then we obtain the result below\footnote{Note that for simplicity we take the two hard scales 
to be equal here, $Q_A^2 = Q_B^2 = Q^2$, and only write down the unpolarised diagonal contribution in colour, flavour 
and spin space. The contributions associated with spin polarisation (longitudinal or transverse) are expected 
to have a similar structure. On the other hand, it is known that the colour and quark number interference contributions 
will contain Sudakov logarithms -- see e.g. \cite{Diehl:2011tt, Diehl:2011yj}.}:
\begin{align} \label{1v2XSec}
\sigma^{D,1v2}_{(A,B)}(s) = & 2\times\dfrac{m}{2}\int dx_1dx_2dy_1dy_2 \hat{\sigma}_{ik \to A}(\hat{s} = x_1y_1s)\hat{\sigma}_{jl \to B}(\hat{s} = x_2y_2s)
\\ \nonumber
&\times \breve{D}^{ij}_p(x_1,x_2;Q^2)  \int \dfrac{d^2\vect{r}}{(2\pi)^2} \Gamma^{kl}_{p,indep}(y_1,y_2,\vect{r};Q^2)
\end{align}

The quantity $\breve{D}^{ij}_p(x_1,x_2;Q^2)$ is the `accumulated sPDF feed' contribution to the dPDF. This evolves
from a zero initial value at a nonperturbative scale $Q_0 \sim \Lambda_{QCD}$ according to the full dDGLAP equation.
$\Gamma^{kl}_{p,indep}(y_1,y_2,\vect{r};Q^2)$ is the `independent branching' 2pGPD. This sums up the effect of independent 
strongly ordered parton emissions from a nonperturbatively generated parton pair. It evolves according to the dDGLAP
equation with the sPDF feed term removed. There is an additional prefactor of $2$ in \eqref{1v2XSec} because there are 
two sets of 1v2 graphs that give equivalent contributions -- in one set the nonperturbatively generated parton pair emerges 
from the `left' proton, whilst in the other it emerges from the `right' proton.

A critical requirement for the derivations of \eqref{2v1XS4} and \eqref{1v2XSec} to be valid is that parton pairs 
connected only via nonperturbative interactions should have an $\vect{r}$-space distribution that is cut off at values 
of order $\Lambda_{QCD}$ (or equivalently a $\vect{b}$-space distribution that is smooth on scales of size $\ll R_p$).
This appears to be a somewhat reasonable requirement -- at a low scale $Q_0 \sim \Lambda_{QCD}$ there is only the scale 
$\Lambda_{QCD}$ available to set the size of the $\vect{r}$ profile for $\Gamma^{kl}_{p,indep}$, and the evolution of 
$\Gamma_{p,indep}$ essentially preserves the transverse profile to higher scales. What is more, such behaviour for 
$\Gamma_{p,indep}$ would appear to be required in order to get the necessary prefactor of order $1/R_p^2$ in the 2v2 
contribution to DPS, which is calculated according to the following expression (for the diagonal unpolarised contribution):
\begin{align} \label{2v2XSec}
\sigma^{D,2v2}_{(A,B)}(s) = & \dfrac{m}{2}\int dx_1dx_2dy_1dy_2 \hat{\sigma}_{ik \to A}(\hat{s} = x_1y_1s)\hat{\sigma}_{jl \to B}(\hat{s} = x_2y_2s)
\\ \nonumber
&\times \int \tfrac{d^2\vect{r}}{(2\pi)^2} \Gamma^{ij}_{p,indep}(x_1,x_2,\vect{r};Q^2) \Gamma^{kl}_{p,indep}(y_1,y_2,\vect{-r};Q^2)
\end{align}

\vspace{-1mm}

Note that the quantity $(2\pi)^{-2}\int d^2\vect{r} \Gamma^{kl}_{p,indep}(y_1,y_2,\vect{r};Q^2)$ in \eqref{1v2XSec}
is equal to $\tilde{\Gamma}^{kl}_{p,indep}(y_1,y_2,\vect{b}=\vect{0};Q^2)$. This appears to indicate that the 2v1 contribution to
DPS probes independent branching 2pGPDs at zero parton separation. In fact, the result \eqref{1v2XSec} corresponds to a 
broad logarithmic integral over values of $\vect{b}^2$ that are $\ll R_p^2$ but $\gg 1/Q^2$. The quantity 
$\tilde{\Gamma}^{kl}_{p,indep}(y_1,y_2,\vect{b}=\vect{0};Q^2)$ appears as a result of our smoothness assumption on 
$\tilde{\Gamma}^{kl}_{p,indep}(y_1,y_2,\vect{b};Q^2)$.

If one assumes that the independent branching 2pGPD can be approximately factorised according to the prescription in
\eqref{2pGPDdecomp2dPDF}, then the contribution to the DPS cross section from 2v1 graphs is similar to that predicted by the dPDF 
framework, albeit with a different associated `$\sigma_{\mathrm{eff}}$'. Indeed we find that $(\sigma_{eff,2v2})^{-1} =
\int d^2\vect{b} [F(\vect{b})]^2$, whilst $(\sigma_{eff,1v2})^{-1} =  F(\vect{b}=\vect{0})$. If one then makes the
further assumption that $F(\vect{b})$ is approximately Gaussian, one finds that each 2v1 contribution to DPS receives a 
factor of 2 enhancement over the 2v2 contribution from the $(\sigma_{\mathrm{eff}})^{-1}$ geometrical prefactor (as is also 
noted in \cite{Blok:2011bu}). One should perhaps not put too much trust into this exact figure, however -- it clearly relies on 
a number of assumptions whose validity is somewhat uncertain.

\vspace{-3mm}

\section{The Total Cross Section for DPS}

If one were to take the suggestions outlined earlier in this report at face value, then one would obtain the 
following expression for (the unpolarised diagonal contribution to) the total LO DPS cross section:
\vspace{-4mm}
\begin{align} \label{DPSmaster}
\sigma^D_{(A,B)}(s) = \sigma^{D,2v2}_{(A,B)}(s) + \sigma^{D,1v2}_{(A,B)}(s)
\end{align}
with $\sigma^{D,1v2}_{(A,B)}(s)$ and $\sigma^{D,2v2}_{(A,B)}(s)$ being given by the expressions \eqref{1v2XSec}
and \eqref{2v2XSec} respectively. This expression shares some common terms with the DPS cross section formulae
proposed in \cite{Blok:2011bu} and \cite{Ryskin:2011kk}. Looking closely at \eqref{DPSmaster} however, one can 
identify a number of concerning issues with regard to this equation, which indicate that modifications to it may be
required in order to correctly describe the DPS cross section.

The first issue is that we were originally expecting to obtain an expression for the DPS cross section 
looking something like \eqref{DPSXsec1}, with the 2pGPDs in these formulae each having an interpretation 
in terms of hadronic operator matrix elements. Our proposed expression \eqref{DPSmaster} deviates somewhat 
in structure from these expectations (in particular, one would struggle to come up with a 
matrix element representation for $\breve{D}^{ij}_p(x_1,x_2;Q^2)$). This feature is related to the fact that
we have entirely removed the `1v1' contribution from the DPS cross section.

The second issue is that there is a rather sharp distinction in \eqref{DPSmaster} between perturbatively
and nonperturbatively generated parton pairs, with the 2pGPD for the latter $\Gamma^{kl}_{p,indep}$ having
a natural width in $\vect{r}$ space of order $\Lambda$. Does there exist some scale at which we can
(approximately) regard all parton pairs in the proton as being `nonperturbatively generated' in this 
sense (as is assumed in \eqref{DPSmaster})? If so, what is the appropriate value for the scale (presumably
it should be rather close to $\Lambda_{QCD}$)?

A final issue is that in the above, we have largely ignored the interesting and potentially 
important interference and correlated parton contributions to DPS catalogued in \cite{Diehl:2011tt, 
Diehl:2011yj}.

In this report, we have shown that the treatment of 1v1 and 2v1 contributions to DPS by the dPDF framework
of Snigirev et al. appears to be unsatisfactory, and presented the results of a calculation that indicates
that we should include a contribution to the DPS cross section from 2v1 graphs, if we include a contribution
from 2v2 graphs. There appear to be some unsatisfactory features in our alternative suggestion for the DPS
cross section \eqref{DPSmaster}, which perhaps indicates that completely removing any contribution from 1v1
graphs from the DPS cross section is not quite the correct prescription.

%
%
%
%


\begin{footnotesize}

\end{footnotesize}


\begin{thebibliography}{99}




\bibitem{DelFabbro:1999tf}
  A.~Del Fabbro and D.~Treleani,
  ``A double parton scattering background to Higgs boson production at the LHC,''
  Phys.\ Rev.\ D {\bf 61} (2000) 077502
  [hep-ph/9911358].


\bibitem{Zinovev:1982be}
  G.~M.~Zinovev, A.~M.~Snigirev and V.~P.~Shelest,
  ``Equations for many parton distributions in Quantum Chromodynamics,''
  Theor.\ Math.\ Phys.\  {\bf 51} (1982) 523
   [Teor.\ Mat.\ Fiz.\  {\bf 51} (1982) 317].


\bibitem{Snigirev:2003cq}
  A.~M.~Snigirev,
  ``Double parton distributions in the leading logarithm approximation of perturbative QCD,''
  Phys.\ Rev.\ D {\bf 68} (2003) 114012
  [hep-ph/0304172].


\bibitem{Nagy:2006xy}
  Z.~Nagy and D.~E.~Soper,
  ``Numerical integration of one-loop Feynman diagrams for N-photon amplitudes,''
  Phys.\ Rev.\ D {\bf 74} (2006) 093006
  [hep-ph/0610028].

\bibitem{Gaunt:2011xd}
  J.~R.~Gaunt and W.~J.~Stirling,
  ``Double parton scattering singularity in one-loop integrals,''
  JHEP {\bf 1106} (2011) 048
  [arXiv:1103.1888 [hep-ph]].

\bibitem{Harindranath:1998pd}
  A.~Harindranath, R.~Kundu and W.~-M.~Zhang,
  ``Deep inelastic structure functions in light front QCD: Radiative corrections,''
  Phys.\ Rev.\ D {\bf 59} (1999) 094013
  [hep-ph/9806221].

\bibitem{Diehl:2011tt}
  M.~Diehl and A.~Schafer,
  ``Theoretical considerations on multiparton interactions in QCD,''
  Phys.\ Lett.\ B {\bf 698} (2011) 389
  [arXiv:1102.3081 [hep-ph]].

\bibitem{Diehl:2011yj}
  M.~Diehl, D.~Ostermeier and A.~Schafer,
  ``Elements of a theory for multiparton interactions in QCD,''
  arXiv:1111.0910 [hep-ph].

\bibitem{Labastida:1984gy}
  J.~M.~F.~Labastida and G.~F.~Sterman,
  ``Inclusive hadron - hadron scattering in the Feynman gauge,''
  Nucl.\ Phys.\ B {\bf 254} (1985) 425.

\bibitem{Blok:2011bu}
  B.~Blok, Y.~Dokshitzer, L.~Frankfurt and M.~Strikman,
  ``pQCD physics of multiparton interactions,''
  arXiv:1106.5533 [hep-ph].

\bibitem{Paver:1982yp}
  N.~Paver and D.~Treleani,
  ``Multi - quark scattering and large P(t) jet production in hadronic collisions,''
  Nuovo Cim.\ A {\bf 70} (1982) 215.

\bibitem{Ryskin:2011kk}
  M.~G.~Ryskin and A.~M.~Snigirev,
  ``A fresh look at double parton scattering,''
  Phys.\ Rev.\ D {\bf 83} (2011) 114047
  [arXiv:1103.3495 [hep-ph]].











\end{thebibliography}
\end{document}